\documentclass[referee]{raa}            

\usepackage{graphicx,times}             
\usepackage{natbib}
\usepackage{amssymb,amsmath}
\bibpunct{(}{)}{;}{a}{}{,}

\usepackage{hyperref}
\hypersetup{colorlinks = true, linkcolor = green, anchorcolor = red, citecolor = blue, filecolor = red, pagecolor = red, urlcolor = red}

\begin{document}

   \title{A Star-based Method for Precise Wavelength Calibration of the Chinese Space Station Telescope (CSST) Slitless Spectroscopic Survey
}

   \volnopage{Vol.0 (20xx) No.0, 000--000}      
   \setcounter{page}{1}          

   \author{Hai-Bo Yuan
      \inst{1}
   \and Ding-Shan Deng
      \inst{1}
   \and Yang Sun
      \inst{1}
   }
 \institute{Department of Astronomy,
   Beijing Normal University, Beijing 100875, China; {\it yuanhb@bnu.edu.cn}\\
\vs\no
   {\small Received~~2020 July 10; accepted~~2020~~August 31}}
   
\abstract{
The Chinese Space Station Telescope (CSST) spectroscopic survey aims to deliver high-quality low-resolution ($R > 200$) slitless spectra
for  hundreds of millions of targets down to a limiting magnitude of about 21 mag, distributed within a large survey area (17500 deg$^2$) and
covering a wide wavelength range (255-1000 nm by 3 bands GU, GV, and GI).
As slitless spectroscopy precludes the usage of wavelength calibration lamps, wavelength calibration is one of the most challenging issues in
the reduction of slitless spectra, yet it plays a key role in measuring precise  radial velocities of stars and redshifts of galaxies.
In this work,  we propose a star-based method that can monitor and correct for possible errors in the CSST wavelength calibration using normal scientific observations, 
taking advantage of the facts that i) there are about ten million stars with reliable radial velocities now available thanks to spectroscopic surveys like LAMOST, 
ii) the large field of view of CSST enables efficient observations of such stars in a short period of time, and iii) radial velocities of such stars can be reliably measured 
using only a narrow segment  of CSST spectra. We demonstrate that it is possible to achieve a wavelength calibration precision of 
a few $\mathrm{km}\,\mathrm{s}^{-1}$ for the GU band, and about 10 to 20 $\mathrm{km}\,\mathrm{s}^{-1}$ for the GV and GI bands, 
with only a few hundred velocity standard stars. Implementations of the method to other surveys are also discussed.  
}

   \authorrunning{H.-B. Yuan, D.-S. Deng \& Y. Sun}            
   \titlerunning{A Star-based Method for Precise Wavelength Calibration of the CSST}  

   \maketitle

%
%
\section{Introduction}           
\label{sect:intro}

Slitless spectroscopy has been a workhorse for astrophysical science since 1960s (Bowen, 1960), due to its unique capability in recording 
enormous numbers of spectra without any artificial pre-selection.  It has been implemented for surveys from the ground, such as the 
Hamburg/ESO objective-prism survey (Wisotzki et al. 1996), the UK Schmidt Telescope (Clowes et al. 1980; Hazard et al. 1986), and 
the Quasars near Quasars survey (Worseck et al. 2008). Due to  the lack of a slit, each pixel receives the full transmission of the dispersing element.
Therefore,  ground-based slitless spectroscopy suffers from the high background, 
Thanks to the very low sky background and very high spatial resolution,
space-based wide-field slitless spectroscopic surveys have become very powerful tools in astronomy \citep{glazebrook_monster_2005}.
A number of surveys (e.g., McCarthy et al. 1999; Gardner et al. 1998; Teplitz et al. 2003; Pirzkal et al. 2004; Malhotra et al. 2005; Momcheva et al. 2016; Pharo et al. 2020) 
have been conducted by different instruments (ACS, NICMOS, WFC3) on broad of the {\it Hubble Space Telescope} 
to study various emission line galaxies and other targets at high redshifts.

In spite of its unique capability, slitless spectroscopy cope with three common issues in terms of their image acquisition and data reduction, including 
how to obtain/construct an image of the field without the dispersing element in place to allow accurate source location,
how to extract spectra from the final image in crowed fields, and how to perform wavelength and flux calibration. The calibration is the most import and also the most challenging part. 

Slitless spectroscopy precludes the usage of wavelength calibration lamps, so that the dispersion solution as a function of the reference position 
must be calibrated in advance, either on-ground or in space, or both. 
Traditionally, on-ground calibrations provide an initial dispersion solution for wavelength calibration.
Then in-orbit calibrations are needed to give an updated and better solution. Such calibrations can only be conducted by observations of specific astrophysical sources.
For this purpose, compact, bright, and  stable targets  in a sparse field and with a good grid of emission lines are typically chosen (see Pasquali et al. 2006).
Such targets include planetary nebulae in external galaxies,  Ae stars, Be stars, Wolf-Rayet stars, cataclysmic variable stars, young stellar objects, and AGNs.
Due to field effects and geometric distortions that cause field-dependent dispersion solution and wavelength zero point, 
calibration targets have to be observed at quite a number of  positions across the field of view in order to map spatial variations in the trace- and wavelength solutions. 
The calibration result is stored and used in the form of a configure file that contains all the instrument dependent parameters.

The Chinese Space Station Telescope (CSST) is a 2 meter space telescope  of a large field of view of 1.1 ${\mathrm{deg}^{2}}$.
It will simultaneously carry out both photometric and slitless grating spectroscopic surveys, covering a large sky area of
17500 ${\mathrm{deg}^{2}}$ but at a high spatial resolution of $\sim 0.15''$ in about ten years (Zhan 2011, 2018; Cao et al. 2018; Gong et al. 2019).
The imaging survey has 7 photometric filters, i.e. $NUV , u, g, r, i, z$, and $y$,  covering 255 --1000 nm from the near-ultraviolet (NUV) to the near infrared (NIR).
While the slitless spectroscopic survey has 3 bands, GU (255-420 nm), GV (400-650 nm), and GI (620-1000 nm).
Complementary to the CSST imaging survey, the CSST slitless spectroscopic survey aims to deliver high quality spectra covering 250 -- 1000 nm at R  larger than 200
for a magnitude-limited sample of hundreds of millions of stars and galaxies. Precise wavelength and flux calibrations are required 
to achieve its various scientific goals, from the nature of dark matter and dark energy, large scale structure and cosmology,  galaxy formation and evolution,
active galactic nucleus (AGNs),  to the Milky Way and near-field cosmology, and stellar physics.
In this work, we focus on the issue of wavelength calibration.

In previous paper (Sun et al. 2020, hereafter Paper I), we show that the CSST slitless spectra have the capability to deliver stellar radial velocities to a precision of $2 - 4 \,\mathrm{km}\,\mathrm{s}^{-1}$ for
AFGKM types of stars at SNR = 100. Considering that accurate radial velocities have been obtained for over ten million of stars by large scale spectroscopic surveys such as 
LAMOST (Zhao et al. 2012),  and upcoming surveys such as  WEAVE (Bonifacio et al. 2016), DESI (DESI Collaboration 2016), SDSS-V (Kollmeier et al. 2017), and 4MOST (de Jong et al. 2019) 
will deliver radial velocities  for more and more stars,
we propose a new star-based method for precise wavelength calibration of the CSST slitless spectroscopic survey.
The method uses enormous numbers of stars (absorption lines rather than emission lines) of known radial velocities observed during normal scientific observations 
as wavelength standards to  monitor and correct for possible errors in wavelength calibration. 
The paper is organized as follows: We introduce our method in Section 2.  We present a simple yet non-trivial verification of our method in Section 3. 
In Section 4, we discuss implications and future improvements of our method. We conclude in Section 5.

\section{Method}
\label{sect:method}

In slitless spectroscopy, the location of an object's spectra on the detector is defined by its position  ($x_{ref}, y_{ref}$)  in the reference frame of the slitless image. 
Therefore,  values of  ($x_{ref}, y_{ref}$) must be firstly determined to set the absolute wavelength scale,. 
It is straightforward in the case of the presence of a direct image, such as the various slitless spectroscopy modes of {\it HST} instruments. 
In the case of the {\it CSST}, where a direct image is not available but a grating is used as disperser, the zeroth order image, when available, could be used to provide the zero point of the wavelength scale. 
When the zeroth order image is not available,  predicted zeroth order image based on astrometric solution, could be used.  
Therefore, in this work, we presume that the zeroth order image ($x_{ref}, y_{ref}$)  is always available.

The wavelength solutions as well as spectral trace are defined with respect to a reference position,  ($x_{ref}, y_{ref}$), and can be described as polynomials.
For example, in the aXe software designed to reduce data from the various slitless spectroscopy modes of {\it HST} instruments  (Kummel et al. 2009), the trace is  describe as: 
\begin{eqnarray}
\Delta y(\Delta x ) = a_0 + a_1 * \Delta x + a_2 * \Delta x^2 + \ldots
\label{eq_trdesc}
\end{eqnarray}
with $(\Delta x, \Delta y) = (x_{trace}-x_{ref}, y_{trace}-y_{ref})$   the offset of the image coordinates ($x_{trace}, y_{trace}$) from ($x_{ref}, y_{ref}$).
Wavelength solutions are described as: 
\begin{eqnarray}
\lambda(l) = l_0 + l_1 * l + l_2 * l^2 + \ldots 
\label{eq_grdisp}
\end{eqnarray}
or
 \begin{eqnarray}
\lambda(l) = l_1 + \frac{l_2}{(l-l_0)} + \frac{l_3}{(l-l_0)^2} + \frac{l_4}{(l-l_0)^3} + \ldots 
\label{eq_prdisp}
\end{eqnarray} 
with  $l$ the distance along the trace. In order to take variations of the trace description and dispersion solution as a function of object position into account,  
 all quantities $a_0, a_1, a_2, \ldots, l_0, l_1, l_2, \ldots$ in Equations (1) -- (3) 
are field-dependent  2D polynomials, and  are functions of ($x_{ref}, y_{ref}$). 
For example, the quantity $a$ given as a 2nd order 2D polynomial  is:
\begin{eqnarray}                                                
a &=& \alpha_0 +  \alpha_1 * x_{ref} + \alpha_2 * y_{ref} + \alpha_3 * x_{ref}^2 + \alpha_4 * x_{ref}*y_{ref} + \alpha_5 * y_{ref}^2.          
\label{eq_fdepend}                                              
\end{eqnarray}

Due to the large field of view of the CSST, spatial variations of the dispersion solution are expected to be strong.
It will be very time-consuming to map out such variations with very fine grids. Temporal variations are also expected. 
To avoid possible large calibration errors with the traditional method, we  propose a new star-based method for precise wavelength calibration of the CSST slitless spectroscopy survey.
The method uses enormous numbers of stars of known radial velocities observed during science observations.
We assume a smooth temporal variations of wavelength solution, i.e., wavelength solution is stable for a period of time, e.g., a few hours. 
We assume that errors in wavelength solution during the period of time can be described as:
 \begin{eqnarray}
\Delta\lambda(l) = l'_0 + l'_1 * l + l'_2 * l^2 + \ldots 
\label{eq_grdisp}
\end{eqnarray}
or 
 \begin{eqnarray}
\Delta\lambda(l) = l'_1 + \frac{l'_2}{(l-l'_0)} + \frac{l'_3}{(l-l'_0)^2} + \frac{l'_4}{(l-l'_0)^3} + \ldots 
\label{eq_prdisp}
\end{eqnarray} 
or more generally, 
 \begin{eqnarray}
\Delta\lambda= f(x_{ref}, y_{ref}, \lambda) = f(x_{ref}, y_{ref}, l) 
\label{eq_grdisp}
\end{eqnarray}
Replacing $l$ by $\lambda$, Equation (5) can be rewritten as:
 \begin{eqnarray}
\Delta\lambda(\lambda) = l'_0 + l'_1 * {\lambda} + l'_2 * {\lambda}^2 + \ldots 
\label{eq_grdisp}
\end{eqnarray}

As in Equations (1) -- (3),  all quantities $l'_0, l'_1, l'_2, \ldots$ in Equations (5) -- (8) are 2D polynomial functions of ($x_{ref}, y_{ref}$). 

During this period of time, a large number ( hundreds or more)  of stars with known radial velocities will be observed and have good SNRs.
 Each GU/GV/GI band spectrum can be divided into several segments. For each segment, we can measure its radial velocity using the original dispersion solution
 and compare it with expected value. The difference in radial velocities can be converted into the offset/error in the  original dispersion solution.
 With hundreds of offsets across the  whole field of view and in different wavelengths, we can map out and then correct for the offsets $\Delta\lambda$ 
 as functions of $x_{ref}, y_{ref},  and\,\lambda$. 
 In this way, we can achieve a better wavelength calibration, without any new cost in observing time.

\section{A simple verification}
\label{sect:R}

\begin{figure}
   \centering
   \includegraphics[width=\textwidth, angle=0]{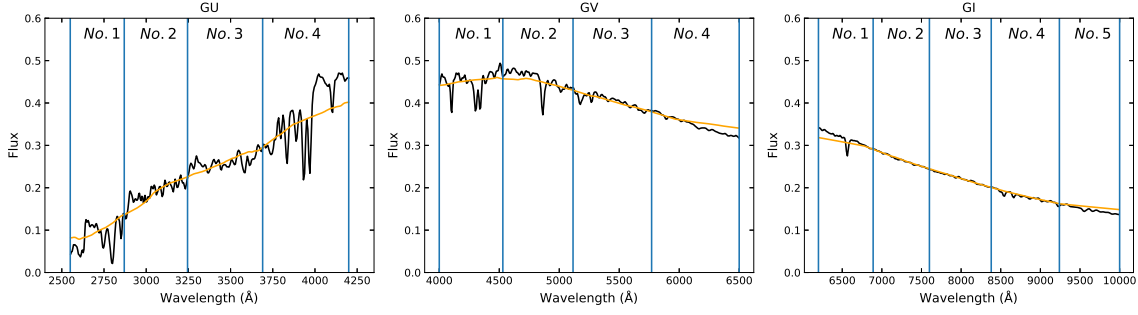}
   \caption{ Example of segment spectrum from $\mathrm{HD 196218}$ ($T_{\mathrm{eff}} = 6207\,$K, $\log \mathrm{g} = 4.11$ dex and $\mathrm{[Fe/H]} = -0.19$ dex). The black and yellow lines are the degraded (R = 250) NGSL spectrum and its  continuum obtained by the simple approach in Paper\,I, respectively. The boundaries  between adjacent segments (marked with serial numbers, $No.i$) are indicated by vertical blue lines.
The wavelengths of these blue lines are 2550, 2870, 3245, 3690, 4200\,\AA\  in the GU band, 4000, 4530, 5115, 5770, 6500\,\AA\  in the GV band, and 6200, 6890, 7600, 8380, 9240, 10000\,\AA\  in the GI band, respectively. Note the prominent metal absorption lines in the GU band spectrum compared to those in the GV and GI bands.}
   \label{Fig1}
\end{figure}

The precision of the new method depends on how many stars can be used  and how precise can we measure stellar radial velocities from their  narrow segment  spectra.
In this section, we first use the same data and method as in Paper I to estimate uncertainties in  measuring radial velocities  using different narrow segment of the CSST spectra.
We use spectra degraded to R = 250 from the  NGSL (Heap \& Lindler 2007; Koleva \& Vazdekis 2012) to simulate the CSS-OS slitless spectroscopic observations.
Spectra normalized  by the simple normalization approach in Paper I,  i.e., a moving average method with window size of 51 pixels, are used.
We split each GU/GV band spectrum into 4 segments and GI band spectrum into 5 segments.  The wavelength ranges of each segment spectrum are shown in Fig.~\ref{Fig1}. 
Then for each segment spectrum, we use the cross-correlation function (CCF) method to measure stellar radial velocities and Monte Carlo simulations to estimate their errors at SNR = 100, as in Paper I. 
Note here the SNR refers to SNR per pixel at a sampling rate of 3 pixels per resolution element.

\begin{figure}
   \centering
   \includegraphics[width=\textwidth, angle=0]{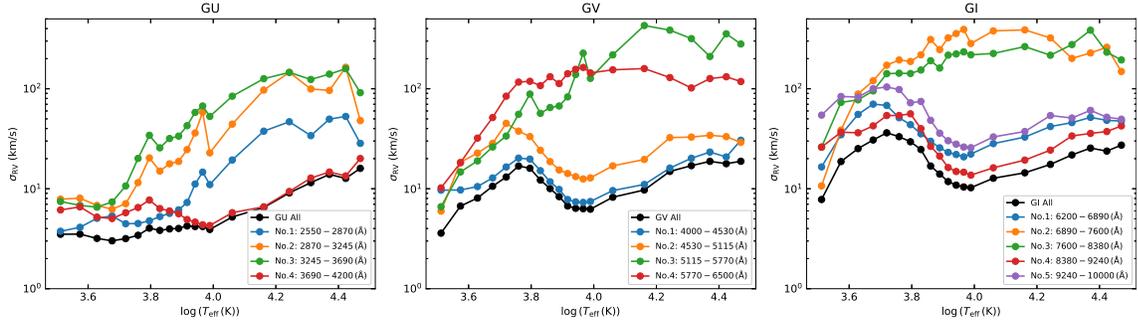}
   \caption{ The median values of $\sigma_{\mathrm{RV}}$ as a function of $T_{\mathrm{eff}}$ at SNR = 100 for different segments. The values for the whole GU/GV/GI band are also plotted for comparison.}
   \label{Fig2}
\end{figure}

We divide the sample into different $T_{\mathrm{eff}}$ bins. The bin size is 500 K for $T_{\mathrm{eff}}$ from 3000 to 10000 K,  and is 3000 K for hotter stars.
Fig.~\ref{Fig2} plots the median values of $\sigma_{\mathrm{RV}}$  as a function of  $T_{\mathrm{eff}}$ at SNR = 100 for different segments of  spectra from the GU (left), GV (middle), and GI (right) bands.
The values for the whole GU/GV/GI band are also plotted for comparison.
The median values of $\sigma_{\mathrm{RV}}$ are also listed in Tab.~\ref{Tab1}.
The results show that the overall trends are consistent with Paper I, but with relatively larger values of $\sigma_{\mathrm{RV}}$  due to narrower wavelength ranges used.

\begin{table}[]
   \bc
   
   \caption[]{The median values of $\sigma_{\mathrm{RV}}$ at SNR = 100 for each segment at different temperature bins.}
   \label{Tab1}
   
   \setlength{\tabcolsep}{3pt}
   \small
   \begin{tabular}{cccccccccccccccccc}
      \hline
      $T_{\mathrm{eff}}$ Range   & \multicolumn{17}{c}{$\sigma_{\mathrm{RV}}$}                                                                 \\
      (K)               & \multicolumn{17}{c}{$(\mathrm{km}\,\mathrm{s}^{-1})$}                                                         \\ \cline{2-18} 
                        & \multicolumn{4}{c}{GU}      &  &  & \multicolumn{4}{c}{GV}      &  &  & \multicolumn{5}{c}{GI}              \\ \cline{2-5} \cline{8-11} \cline{14-18} 
                        & No.1 & No.2  & No.3  & No.4 &  &  & No.1 & No.2 & No.3  & No.4  &  &  & No.1 & No.2  & No.3  & No.4 & No.5  \\ \hline
      {[}3000,3500)     & 3.8  & 7.9   & 7.5   & 6.1  &  &  & 9.7  & 6.0  & 6.6   & 10.2  &  &  & 16.5 & 10.7  & 26.1  & 26.1 & 54.5  \\
      {[}3500,4000)     & 4.1  & 8.0   & 6.8   & 6.6  &  &  & 9.7  & 18.4 & 14.7  & 18.2  &  &  & 34.7 & 38.6  & 73.1  & 36.8 & 83.7  \\
      {[}4000,4500)     & 5.1  & 6.8   & 6.5   & 5.2  &  &  & 10.5 & 22.7 & 19.0  & 32.1  &  &  & 55.4 & 88.6  & 77.5  & 36.3 & 82.6  \\
      {[}4500,5000)     & 5.4  & 6.2   & 7.4   & 5.0  &  &  & 12.8 & 28.5 & 26.2  & 51.6  &  &  & 70.2 & 120.3 & 95.2  & 42.6 & 100.2 \\
      {[}5000,5500)     & 4.5  & 7.1   & 10.7  & 5.8  &  &  & 16.5 & 45.1 & 33.7  & 84.3  &  &  & 68.1 & 172.0 & 141.6 & 54.0 & 104.2 \\
      {[}5500,6000)     & 4.5  & 11.5  & 20.1  & 6.5  &  &  & 20.2 & 37.7 & 55.6  & 116.7 &  &  & 51.3 & 194.2 & 143.1 & 54.1 & 98.2  \\
      {[}6000,6500)     & 4.8  & 20.3  & 34.2  & 7.7  &  &  & 19.8 & 33.2 & 88.6  & 118.8 &  &  & 43.7 & 187.5 & 142.1 & 56.2 & 72.4  \\
      {[}6500,7000)     & 5.3  & 15.0  & 25.7  & 6.3  &  &  & 15.2 & 24.2 & 56.8  & 107.5 &  &  & 35.4 & 218.4 & 154.0 & 39.9 & 74.7  \\
      {[}7000,7500)     & 5.7  & 17.8  & 31.8  & 6.0  &  &  & 11.7 & 18.7 & 64.9  & 132.0 &  &  & 29.9 & 311.2 & 190.9 & 26.5 & 48.4  \\
      {[}7500,8000)     & 6.1  & 18.7  & 33.6  & 5.6  &  &  & 9.8  & 15.4 & 67.3  & 112.7 &  &  & 25.7 & 246.6 & 161.2 & 21.2 & 36.0  \\
      {[}8000,8500)     & 7.3  & 24.6  & 42.7  & 4.9  &  &  & 7.8  & 14.2 & 83.0  & 141.5 &  &  & 23.1 & 323.6 & 217.0 & 16.3 & 30.2  \\
      {[}8500,9000)     & 11.2 & 36.1  & 58.1  & 4.6  &  &  & 7.4  & 13.2 & 138.8 & 156.4 &  &  & 22.0 & 356.7 & 222.9 & 14.9 & 27.9  \\
      {[}9000,9500)     & 14.7 & 58.3  & 67.3  & 4.4  &  &  & 7.3  & 12.5 & 227.2 & 163.5 &  &  & 20.9 & 392.3 & 233.9 & 14.6 & 26.1  \\
      {[}9500,10000)    & 11.0 & 22.9  & 52.9  & 4.3  &  &  & 7.5  & 12.8 & 127.0 & 144.1 &  &  & 22.2 & 283.6 & 218.7 & 13.7 & 25.7  \\
      {[}10000,13000)   & 19.4 & 44.3  & 84.3  & 5.8  &  &  & 9.6  & 16.9 & 218.4 & 154.7 &  &  & 28.3 & 379.4 & 225.5 & 16.2 & 33.0  \\
      {[}13000,16000)   & 37.7 & 97.1  & 126.1 & 6.6  &  &  & 11.0 & 19.6 & 429.5 & 159.2 &  &  & 32.9 & 388.0 & 264.2 & 19.3 & 37.4  \\
      {[}16000,19000)   & 46.8 & 146.0 & 145.8 & 9.4  &  &  & 16.1 & 32.5 & 385.9 & 129.5 &  &  & 41.9 & 322.5 & 216.5 & 24.3 & 54.0  \\
      {[}19000,22000)   & 34.0 & 99.5  & 123.9 & 12.8 &  &  & 20.2 & 32.9 & 317.5 & 101.9 &  &  & 45.8 & 201.3 & 276.8 & 33.7 & 51.0  \\
      {[}22000,25000)   & 49.7 & 96.3  & 140.2 & 14.7 &  &  & 23.3 & 34.3 & 210.3 & 126.6 &  &  & 51.7 & 228.1 & 385.5 & 35.8 & 60.8  \\
      {[}25000,28000)   & 52.9 & 164.0 & 158.5 & 13.4 &  &  & 20.8 & 33.2 & 354.9 & 131.9 &  &  & 48.4 & 260.1 & 232.2 & 37.3 & 52.0  \\
      {[}28000,31000{]} & 28.5 & 48.1  & 91.4  & 20.1 &  &  & 30.6 & 29.1 & 280.5 & 118.1 &  &  & 47.5 & 148.9 & 194.3 & 42.5 & 49.6  \\ \hline
   \end{tabular}
   \ec
   \end{table}

To further verify our method, we have performed Monte Carlo simulations to estimate calibration errors of the method with different number of stars used. 
We assume that the errors of original wavelength calibration of CSST can be described by Equation (8) at order 2.
$l'_0, l'_1, l'_2$  coefficients in  Equation (8) are 2nd order  2D polynomial functions of ($x_{ref}, y_{ref}$), as given by Equation (4).
Therefore, a total number of 18 free parameters are needed to describe calibration errors. 
We assume a perfect wavelength calibration, i.e., all the 18 parameters are zero. 
Then we use $N$ stars randomly selected from the LAMOST DR5 (Luo et al. 2015), corresponding to $4\times N$ ($5\times N$ for the GI band) measurements of calibration errors at 4 (5 for the GI band) given wavelengths but different  ($x_{ref}, y_{ref}$) positions, to constrain  the 18 free parameters. 
The stars are assumed to be uniformly distributed in the field of view, i.e.,  both $x_{ref}$ and $y_{ref}$ values are uniformly distributed between 0 and 1.
We also assume that all these stars are well observed by the CSST and have SNRs of 100 and their velocity errors are given by Tab.~\ref{Tab1} according to their temperatures.
With newly derived 18 parameters, predicted $\Delta\lambda(\lambda)$ values at 10000 positions that are uniformly distributed in the field of view are computed and compared to the assumed values of zero. 
The precision is indicated by the dispersions of $\Delta\lambda(\lambda)$. N = 100, 400 are used. 

\begin{figure}
   \centering
   \includegraphics[width=\textwidth, angle=0]{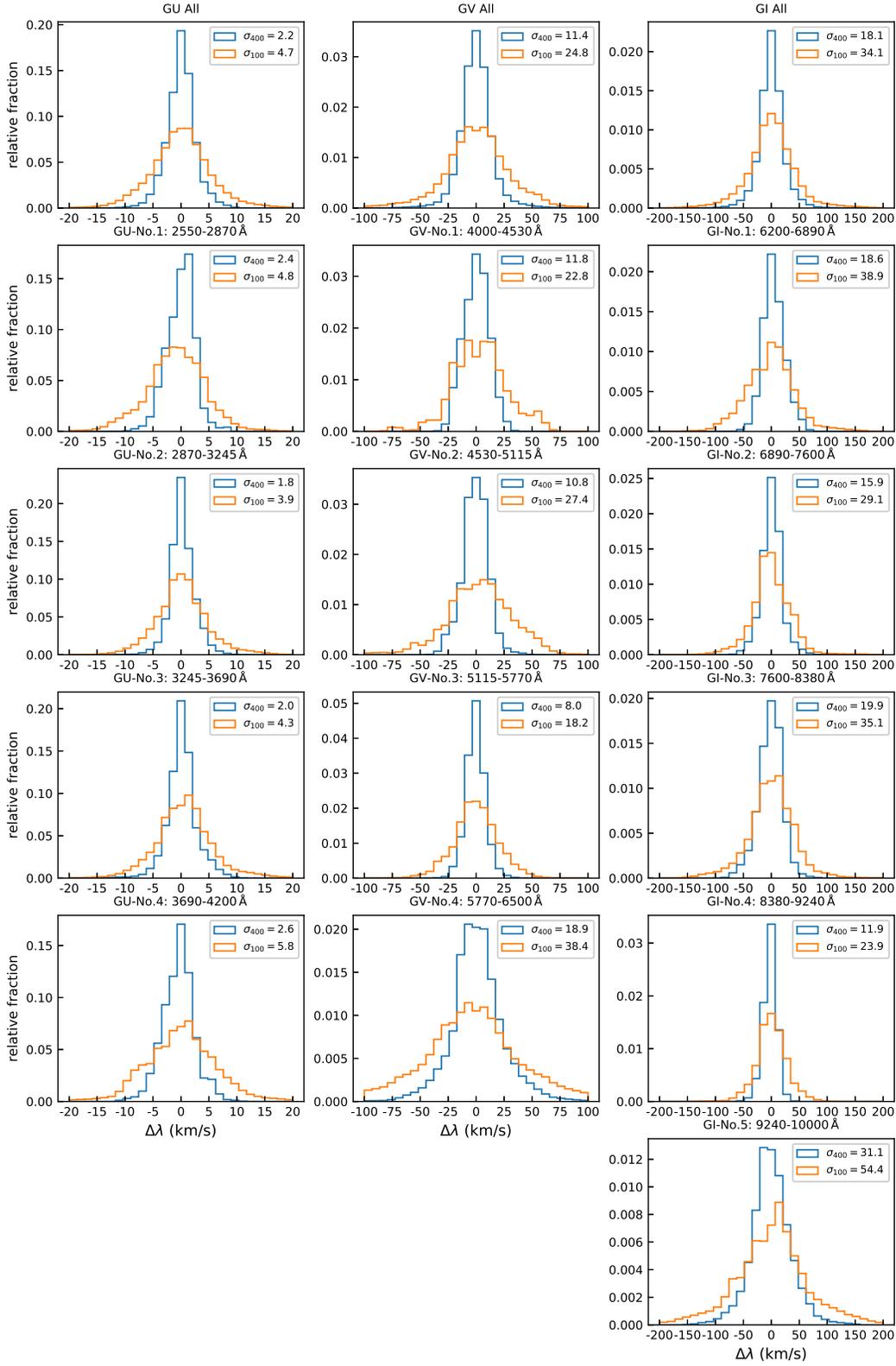}
   \caption{Distribution of wavelength calibration errors for the GU (left), GV (middle), and GI (right) bands. 
   The results with N = 100 and 400 are denoted by blue and orange lines, respectively. The dispersion values are also labeled.}
   \label{Fig3}
\end{figure}

\begin{figure}
   \centering
   \includegraphics[width=\textwidth, angle=0]{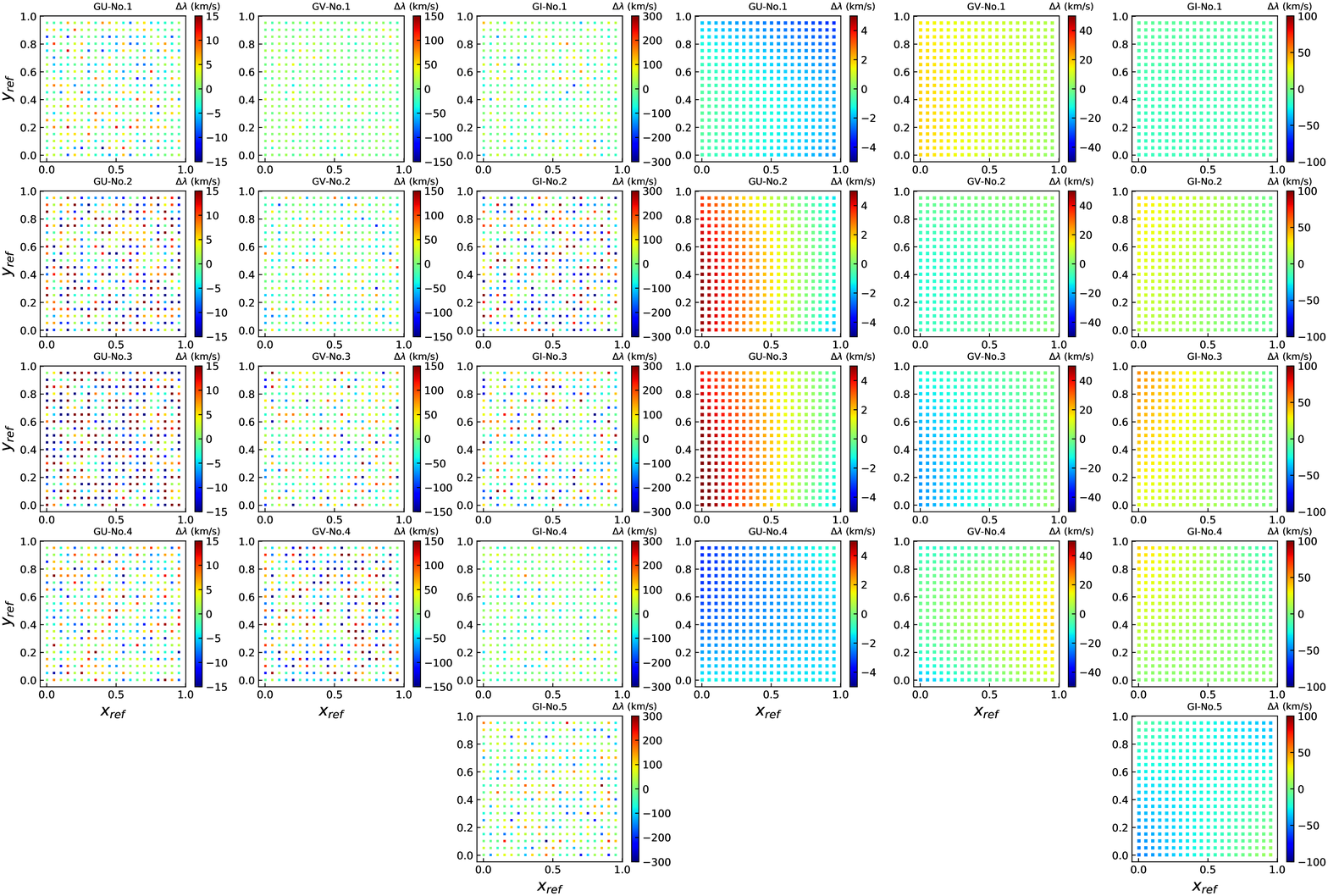}
   \caption{ Left: Spatial distributions of simulated $\Delta\lambda(x_{ref}, y_{ref}, \lambda)$ values from a typical simulation with N = 400. Right: Spatial distributions of calibration errors from the same simulation. }
   \label{Fig4} 
\end{figure}

Fig.~\ref{Fig3} shows the distribution of wavelength calibration errors 
using a total number of 100 simulations. 
The left, middle, and right panels show results for the GU, GV and GI bands, respectively.
The dispersion values are labeled in Fig.~\ref{Fig3} and listed in Tab.~\ref{Tab2}.
It can be seen that with 400 stars, the method can achieve a wavelength calibration precision of about 2, 10, and 20 $\mathrm{km}\,\mathrm{s}^{-1}$ for the GU, GV and GI bands, respectively.
With 100 stars, the numbers are increased by a factor of about two, as expected.
Fig.~\ref{Fig4} shows the spatial distributions of both the simulated  and predicted calibration errors $\Delta\lambda(x_{ref}, y_{ref}, \lambda)$  from a typical simulation at N = 400.
Note that systematic errors in radial velocity measurements are much smaller than wavelength calibration errors, due to the usage of the full range of spectra.

\begin{table}[]
   \bc
   \caption[]{Wavelength calibration errors for the GU/GV/GI bands at N=100, 400.}
   \label{Tab2}

   \setlength{\tabcolsep}{3pt}
   \small
   \begin{tabular}{ccc}
   \hline
   Wavelength Range    & \multicolumn{2}{c}{Dispersions of $\Delta\lambda$}   \\
   (\AA)  & \multicolumn{2}{c}{$(\mathrm{km}\,\mathrm{s}^{-1})$} \\ \cline{2-3} 
   & N = 100                   & N = 400                  \\ \hline
GU-All: 2550-4200   & 4.7  & 2.2  \\
GU-No.1: 2550-2870  & 4.8  & 2.4  \\
GU-No.2: 2870-3245  & 3.9  & 1.8  \\
GU-No.3: 3245-3690  & 4.3  & 2.0  \\
GU-No.4: 3690-4200  & 5.8  & 2.6  \\ \hline
GV-All: 4000-6500   & 24.8 & 11.4 \\
GV-No.1: 4000-4530  & 22.8 & 11.8 \\
GV-No.2: 4530-5115  & 27.4 & 10.8 \\
GV-No.3: 5115-5770  & 18.2 & 8.0  \\
GV-No.4: 5770-6500  & 38.4 & 18.9 \\ \hline
GI-All: 6200-10000  & 34.1 & 18.1 \\
GI-No.1: 6200-6890  & 38.9 & 18.6 \\
GI-No.2: 6890-7600  & 29.1 & 15.9 \\
GI-No.3: 7600-8380  & 35.1 & 19.9 \\
GI-No.4: 8380-9240  & 23.9 & 11.9 \\
GI-No.5: 9240-10000 & 54.4 & 31.1 \\ \hline
   \end{tabular}
   \ec
   \end{table}

\section{Discussion}
\label{sect:D}

The results show that the star-based method has the potential to achieve a wavelength calibration precision of about a few $\mathrm{km}\,\mathrm{s}^{-1}$ for the GU band, about 10
$\mathrm{km}\,\mathrm{s}^{-1}$ for the GV band, and about  20 $\mathrm{km}\,\mathrm{s}^{-1}$ for the GI band, with only a few hundred stars.
Given the capabilities of CSST spectra in measuring stellar radial velocities as demonstrated in Paper\,I,  it suggests that  the CSST spectroscopic survey is very promising to 
deliver reliable velocities (with uncertainties about 10 $\mathrm{km}\,\mathrm{s}^{-1}$) for a unique magnitude-limited sample of stars  with huge numbers for Galactic and stellar sciences, 
such as Galactic kinematics and  searching for high-velocity stars and hyper-velocity stars (Brown 2015).

A few hundred well-observed velocity standard stars by the CSST are required to make the method work well. 
The CSST slitless spectroscopic survey can reach about 17 mag at SNR = 100.
It means that most LAMOST targets ($r < 17.8$ mag) can serve as good velocity standard stars.
Upcoming surveys such as  WEAVE (Bonifacio et al. 2016), DESI (DESI Collaboration 2016), SDSS-V (Kollmeier et al. 2017), and 4MOST (de Jong et al. 2019),  will 
also provide millions of  high-quality velocity standard stars. Given the field of view for each CSST grating of about 100 arcmin$^2$, the exposure time of 150s, and the typical number 
density of velocity standard stars of a few hundred per deg$^2$, 
we expect that it will take a few hours to collect spectra of a few hundred velocity standard stars.  As long as wavelength solution is stable for a few hours or longer, the method should work well.
In cases of very unstable wavelength solution, a number of specified wavelength calibration fields,  which have a high density (thousands per deg$^2$) of velocity standard stars,                                   
 can be designed in advance to increase the number of velocity standard stars to be used in a very short time.

Although we focus on normal stars in this work, we note that any point sources (Ae stars, Be stars, Wolf-Rayet stars, cataclysmic variable stars, young stellar objects, and AGNs) 
with well detected emission lines and known velocities 
can also be directly used  in the framework of the new method. If necessary, compact galaxies will well measured redshifts can also be included.
We ignore velocities variations caused by binary stars. However, only a very small fraction of binary stars show velocity variations 
larger than 10 $\mathrm{km}\,\mathrm{s}^{-1}$ (e.g., Tian et al. 2018). 
In this work, we also ignore uncertainties in the $x_{ref}$ and  $y_{ref}$ values of velocity standard stars. 
Their uncertainties are expected to be very small as most velocity standard stars are very bright and will have high quality zero-order images and/or accurate positions by Gaia (Lindegren et al. 2018). 
We also ignore velocity uncertainties in the velocity standard stars,  considering that their typical errors (a few $\mathrm{km}\,\mathrm{s}^{-1}$) are comparable to or smaller than the numbers in Table\,1.

In this work, we have assumed that each narrow segment spectrum suffers the same systematic errors in wavelength calibration. 
In real cases, it is likely that wavelength calibration errors depend on wavelength even for a very narrow wavelength range.  This problem can be solved in 
a very straightforward way by using pixel-based optimizations, which will be implemented in the future.

The method can not only applied to the CSST, but also other spectroscopic surveys. 
We note that velocity uncertainties of most spectroscopic surveys such as LAMOST, SDSS, and APOGEE (Huang et al, to be submitted), 
are dominated by systematic errors due to wavelength calibration errors.
The method can be applied to these surveys to investigate and then correct for sources of wavelength-dependent wavelength calibration errors.

\section{Summary}
\label{sect:Sum}

Wavelength calibration plays a key role in measuring precise  radial velocities of stars and redshifts of galaxies
for the CSST slitless spectroscopic survey, yet it is one of the hardest issues in the data reduction process due to 
the failure of  wavelength calibration lamps.

In this work, considering the facts that  i) there are about ten million stars with reliable radial velocities now available thanks to spectroscopic surveys like LAMOST, 
ii) the large field of view of CSST enables efficient observations of such stars to a large number and in a short period of time, and iii) radial velocities of such stars can be reliably measured using only a narrow segment  of CSST spectra, 
we propose a star-based method that can monitor and correct for possible errors in the CSST wavelength calibration using normal scientific observations.
With a simple simulation, we demonstrate that it is possible to achieve a wavelength calibration precision
of bout a few $\mathrm{km}\,\mathrm{s}^{-1}$ for the GU band, about 10
$\mathrm{km}\,\mathrm{s}^{-1}$ for the GV band, and about  20 $\mathrm{km}\,\mathrm{s}^{-1}$ for the GI band, with only a few hundred stars.
Given the possible high precision of wavelength calibration and radial velocities, the CSST spectroscopic survey can deliver unique samples and enable interesting science such as Galactic kinematics 
and searching for hyper-velocity stars. The method can also be applied to other spectroscopic surveys such as LAMOST.

\normalem
\begin{acknowledgements}
Dedicated to the Department of Astronomy of Beijing Normal University, the 2nd astronomy program in the modern history of China. 
This work is supported by the National Key Basic R\&D Program of China via 2019YFA0405500, National Natural Science Foundation of China through the project NSFC 11603002,
 and Beijing Normal University grant No. 310232102.
 \end{acknowledgements}

\bibliographystyle{raa}
\bibliography{WaveCalib.bib}

\clearpage

\end{document}